\documentstyle[11pt,newpasp,twoside,epsf]{article}
\begin{document}

\title{REM -- Rapid Eye Mount. A fast slewing robotized infrared telescope}
\author{Stefano Covino$^1$, Filippo Zerbi$^1$, Guido Chincarini$^1$, Marcello
Rodon\'o$^2$, Gabriele Ghisellini$^1$, Angelo Antonelli$^3$, Paolo Conconi$^1$,
Giuseppe Cutispoto$^2$, Emilio Molinari$^1$... {\em on behalf of the REM team}}
\affil{$^1$Oss. Astronomico di Brera, Via Bianchi 46, 23807 Merate
(LC), Italy}
\affil{$^2$Oss. Astrofisico di Catania, Via S. Sofia 78, 95123
Catania, Italy}
\affil{$^3$Oss. Astronomico di Roma, Via Frascati 33, 00044 Rome,
Italy}

\begin{abstract}
The main goal of the REM project is the observation of prompt afterglow of
Gamma--Ray Burst (GRB)  events. Such observations at Near InfraRed (NIR)
wavelengths are even very promising, since they allow to monitor high--z
Ly-$\alpha$ absorbed bursts as well as events occurring in dusty star--forming
regions. In addition to GRB science, a large amount of time ($\sim 40\%$)
will be available for different scientific targets: among these the study of
variability of stellar objects open exciting new perspectives.
\end{abstract}

\section{Introduction}

In the last decade, mainly due to the observations by {\it Beppo}SAX and related
optical Earth--based follow up, the physics of GRBs and their afterglow started
to be understood. However, the intrinsic duration of the observational process,
mainly due to the use of device non--optimized for fast response, did not allow
up to now to monitor the prompt afterglow, i.e. the emission in the few
seconds after the burst (except in one case, GRB\,990123). In the next few
years, dedicated missions such as HETEII or Swift and of
other space--borne high--energy observatories with trigger capabilities
(INTEGRAL, AGILE, etc.) will allow in a few years to accumulate a statistical
significant number of GRB detections.

In this context, REM will be operational at the European Southern Observatory
of La Silla within October 2002. REM shares with other telescopes the
characteristics of fast pointing and full robotisation but it has the unique
feature to be equipped with a high throughput NIR Camera. Any possible
optimization of the trigger sources will leave some time in which REM can not
observe any GRB afterglow due to latitude/longitude constraint. We estimate such
{\it Idle Time} to be of the order of $40 \%$ of REM observing time. During the
idle time REM will be used for multifrequency monitoring of variable objects.
Three {\it Key programmes} have been identified by the REM science team as
particularly suitable for REM: {\bf a}) monitoring of Blazars, {\bf b})
monitoring of flare and variable stars and {\bf c}) observation of IR
counterpart of galactic Black Hole (BH) candidates.

\section{REM telescope and instrumentation}

REM has a classical Ritchey--Chretien optical scheme mounted in an
alt--azimu\-thal configuration with two Nasmyth focal stations, a 60\,cm primary
mirror and a total focal ratio of F/8. Such optical figures allow a compact
structure that provides the needed stiffness for fast motion and windy
environmental conditions such as those expected in a fully deployable dome at
the La Silla observatory.

One of the Nasmyth focal station will be equipped with a fully cryogenic NIR
($0.9 - 2.3\,\mu m$) camera. The camera design had been developed with high
throughput as major goal. The camera has a focal reducer scheme and is made of
two detached doublets of Silica and CaF$_2$ and further Silica correction lens.
The pupil of the system is re--formed between the doublets allowing to locate a
Lyot cold stop, imaging filters and grisms for low--resolution slit--less
spectroscopy. The camera has quasi diffraction--limited Optical Quality.
The dithering, needed for IR image processing is obtained via a wobbling wedge
that shifts the image of about 20 pixels in any radial direction from the array
center (Conconi et al. 2001).

The camera will be equipped with a 512x512 Rockwell HAWAII LPE
HgCdTe chip (18\,$\mu$m pitch) with a peak efficiency of 64\% and
values never lower than 56\% between 1 and 2.5\,$\mu$m. The camera
(telescope and filter excluded) is expected to have a transmission
of $T>53$\%. With a scale of 1.16\,arcsec/pixel the camera covers a
FOV of 9.8x9.8\,arcmin$^2$.
According to simulations such an high throughput allows to reach the limit
magnitudes reported in Table\,\ref{tab:magn}.

\begin{table}
\caption{Limit Magnitudes for different S/N and integration time as computed via
the REM telescope simulator.}
\begin{center}
\begin{scriptsize}
\begin{tabular}{||l ||c| c| c| c| c| c| c| c|| }
\hline T int.&Z&Z&J&J&H&H&K&K\\ &S/N=10 &S/N=5 & S/N=10 &S/N=5
&S/N=10 &S/N=5 & S/N=10 &S/N=5\\ \hline 5 sec &17.0&17.7& 15.7
&16.5 & 14.5 &15.3& 13.2& 14.0
\\
30 sec &19.9&20.7& 16.6 &17.4 & 15.5 &16.2& 14.2& 14.9 \\ 600 sec
&24.5&25.3&17.6  &18.3 & 16.7 &17.4& 15.5& 16.3 \\ \hline
\end{tabular}
\end{scriptsize}
\end{center}
\label{tab:magn}
\end{table}

\section{Observation of stellar objects}

In its essence the REM telescope is a very efficient simultaneous ZJHK
photometric device. The plate--scale chosen makes this telescope of limited use
for extended objects or crowded fields but it certainly allow to collect
photometry of point--like source and is therefore ideal to observe stellar
objects in and outside our galaxy.

During the observation of a GRB REM will acquire continuously frames measuring
NIR colors of all the stellar objects contained in such a frame. Among this
object, any variability with time--scale smaller or comparable with the time in
which REM will insist on the field will be detected. This will allows to discard
variable objects from the list of stars to be used for calibration but it will
also allow to discover new variables. Moreover also the colors of the constant
stars will be measured with very high precision.

It is however during the idle time when REM will be a unique opportunity for
many programs of paramount interest for various branches of stellar
astrophysics. Indeed the expected limiting magnitudes of the REM instrument will
easily allow to measure a plethora of classes of known variable stars with
previously un--conceivable efficiency and precision.

In the following section we present a non-exhaustive description of some of the
main fields in stellar astrophysics that have given proof in the recent past to
gain a lot from good quality observations in the NIR.

\subsection{Cepheids and RR Lyrae stars}

Cepheids and RR Lyrae pulsating variables are known to have paramount
astrophysical importance related to their use as distance and age indicators at
the cosmological level. Such an importance is
of course enhanced if we observe at IR wavelengths since the interstellar
absorption in the K-band is about ten times smaller than the one in the
V-band.

A great gain is also attainable by observing Cepheids and RR Lyrae at NIR
wavelength due to the different view of the pulsation given at these
and at optical wavelengths. This
because the light variation depends both on the radius variation and the surface
brightness (or effective temperature) variation but with different weight
depending on the wavelength. Another advantage in IR observation of Cepheids regards the
blanketing. In the 1--3\,$\mu$m region, where the
density of metallic absorption lines is low, this effect is
much reduced

The expected performances of the REM telescope allow the
observation of the largest number of known Cepheids and RR\,Lyrae
stars ever observed systematically in Z, J, H and K so far. Very
bright Cepheids will be within the limiting magnitude of the REM
instrument even in nearby galaxies. Beside gaining a new insight
in the pulsation mechanisms ongoing in these stars we will then
have the opportunity to put a further constraint to major distance
and age calibrations via the application of independent methods
based solely on IR measurements.

\subsection{T--Tauri stars}

T--Tauri stars are low-mass pre--ZAMS stars that were first
distinguished due to their photometric variations.
Monitoring of the variability of T--Tauri stars at visible and NIR wavelength
suggested several causes for such variability derived basically from a picture
of the circumstellar environment in which surrounding material in an
infalling envelope (e.g. Calvet et al 1994) falls onto and
dissipates angular momentum in a circumstellar disk. The presence
of magnetic fields produce cool sun--like spots on the surface of
the star but also channel the disk material on to the stellar
surface producing hot isolated spots (Shu et al 1994).

In the above framework the variability of T--Tauri stars has
multiple causes and various characteristics. The cool starspots
modulate periodically the brightness of the star with amplitude
slowly changing in time--scales of months or years as the spots
evolve. Typical life--time of these spots is of the order
of days or weeks.

On top of these some environmental characteristics such as the
changes in extinction as infalling or orbiting material intersects
the line--of--sight, can cause variability. Extinction variations
can have amplitudes limited only by the optical depth of the
interceding material. The duration of the event depends on the
size of the cloud and its velocity. Such clouds may either
originate in the circumstellar environment or represent pristine
infalling material.

Since the typical amplitude of variation of these stars is between some
tenth to some magnitudes REM represents a valid tool to monitor
these stars in ZJHK with good coverage and S/N ratio: most of the
above described time-scales of variation are as well compatible
with REM observing strategy.

\subsection{ Magnetic activity in late--type stars}

In addition to cool spots, intense flares are the most remarkable manifestation of stellar magnetic
activity. Flare events occur in the atmospheres of several type of stars, from
pre--MS to post--MS cool stars and involve the whole atmosphere, from the
photosphere up to the corona. Due to the rather different physical
characteristics of these different layers, with temperatures and densities
spanning over several decades, flare flux affects a wide range of wavelengths,
from microwaves to X--rays, and possibly Gamma--Rays. Stellar flares are quite
fast with typical time scales of flux increase to the maximum of the order of
10--100\,s and decreases to the pre--flare level on times 10--100 longer (see
Rodon\'o 1990): a striking case of a multi--wavelength phenomenon that, owing to
its fast development, requires really simultaneous observations.

Multi--wavelength studies (see e.g., Rodon\'o et at. 1989)  have shown that on
the occasion of intense flares in the microwave, UV and optical wavebands, the
IR flux decreases. Such negative flares are predicted by a non--thermal model
based on inverse Compton effects by fast electron (Gurzadian 1980) and a thermal
model based on the increased opacity of the $H-$ ion (Grinin 1976). The missing
energy in the K band alone can account for the energy flux increase at all other
wavelengths. Multi-wavelength monitoring of stellar flares are therefore
relevant for the interpretation of magnetic phenomena.

The possibility of GRBs associated with stellar flares has
been predicted, among others, by Becker and Morrison (1974), but no observations
have been devoted to this aspect. Outside of the time devoted to the principal
research objective, namely the GRB afterglow observations in IR, we intend to
use the REM telescope to study negative flares in IR, as well as to ascertain
whether any of the GRB detected by Swift  is associated with a stellar flare
rather than to an extragalactic phenomenon.

Finally, huge flux excess in the far IR have been detected on quiescent dMe active stars with ISO
(Rodon\'o et al. 1999,  Leto et al. 2001)  and it will be interesting to relate
those enhanced IR emissions with REM data.

\subsection{Cataclysmic Variables, Low and High Mass X--Ray Binaries}

Cataclysimic Variables (CVs) are semi--detached binary systems made up
of a White Dwarf (WD) primary and a Main Sequence (MS) secondary. Material
donated by the secondary star is funneled through the L1 Lagrangian point.

The NIR emission in CVs is produced by a variety of components:
{\bf 1}) the accretion disk in non--magnetic systems or the accretion
stream in magnetic systems, {\bf 2}) the accretion region near the
surface of the WD, {\bf 3}) the accretion disk hot spot and
{\bf 4}) the secondary star. Therefore, even photometrically, the NIR
wavelengths offer a plethora of information about the component
stars and the accretion process within CVs. Indeed the classical
approach to constraint SED (Spectral Energy Distribution) models
for CVs is to disentangle within time-resolved NIR-Curves each
component of the emission.

By browsing the catalogue of Cataclysmic variable by Ritter and
Kolb (1998) we find a consistent number of CVs with $V$ magnitude
between 10 and 15. Very little is known about the infrared colours
of these stars. However a number of these objects have been
observed in these band providing evidence of  $K$ band flux larger
than $V$ band flux. The typical amplitude of variations in
the NIR--curves is of the order of $0^m.3$, well within the expected
capabilities of the REM telescope.

Low Mass X--Ray Binaries (LMXRBs) cointaining either a neutron
star or a black hole, accrete matter from the late--type companion
by means of an accretion disk. LMXRBs can be divided into
subclasses according to their location in the Galaxy: bright
Galactic Bulge Sources located near the Galactic Center and
the other LMXRBs in the Galactic disk. The study of the first
subclass has been severely hampered by the heavy obscuration,
which made even the recognistion of the optical counterpart
difficult. The dichotomy between these two LMXRB subclasses is
poorly understood. The IR  band provides us with an ideal
window for observing these systems. REM observation can help
studying the variability of these objects in connection with their X--ray fluxes.

Infrared monitoring of High mass X--ray Binaries (HMXRBs) is
especially important in the case of Be companion stars. These
systems are usually transient: large X--ray outbursts are
associated to shell ejection episodes whereas smaller ones with
passeages of the neutron star companion near periastron. IR
monitoring therefore provides a unique opportunity to test the
environment in which the neutron star is accreting (Negueruela 1998). IR
monitoring of these sources, which are usually bright and variable (up to two
magnitudes), in conjunction with X--ray monitoring can shed light on the
outburst onset mechanism (e.g. propeller vs. direct accretion)  as well as lead
to the prediction for the occurence of X--ray outbursts which could then be
monitored much more accurately, with pointing X--ray instruments.

\end{document}